\documentclass[a4papper,12pt]{article}
\title{Around the goal: Examining the effect of the first goal on the second goal in soccer using survival analysis methods\\}
\author{Daniel Nevo and Ya'acov Ritov}
\usepackage{amsmath,dcolumn}
\usepackage{amsthm}
\usepackage{graphicx}
\usepackage{subfig}
\usepackage[normalem]{ulem}
\usepackage{color}
\usepackage[round]{natbib}
\newtheorem{remark}{Remark.}
\begin{document}
\maketitle

\section*{Abstract}
In this paper we apply survival techniques to soccer data, treating a goal scoring as the event of interest. It specifically concerns the relationship between the time of the first goal in the game and the time of the second goal. In order to do so, the relevant survival analysis concepts are readjusted to fit the problem and a Cox model  is developed for the hazard function. Attributes such as time dependent covariates and a  frailty term are also being considered. We also use a reliable propensity score to summarize the pre-game covariates. The conclusions derived from the results are that a first goal occurrence could either expedite or impede the next goal scoring, depending on the time it was scored. Moreover, once a goal is scored, another goal scoring become more and more likely as the game progresses. Furthermore, the first goal effect is the same whether the goal was scored or conceded.

\section{Introduction}
\label{sintro}
Various models were suggested to describe an assortment of soccer related phenomenons. \citet{Maher} suggested a simple two independent Poisson variables model to the number of goals for each team with four parameters for each team; home attack skills; home defence skills; away attack skills; away defence skills. He also considered a bivariate Poisson distribution and found the value 0.2 to be most appropriate for the correlation. The dependency between the number of goals scored by the two teams was discussed by \citet{DixonColes}. They concluded that this dependency exists mainly in low scoring games (i.e., games in which both teams did not score more than one goal). Their model is based upon Maher's (1982) model, taking into care each team attack and defence skill as well as the home advantage and the correlation mentioned above. They also suggested a way to handle the team ability changes over time and used pseudo likelihood for each time to derive the parameters of the model. Their model was aimed at exploiting possible inefficiencies in the English soccer betting market.

\citet{DixonRobinson} assumed an exponential distribution for the time until the next goal was scored, and examined a two dimensional birth process model with respect to each teams' skills, as well as home advantage and time dependent changes in the score rates (i.e., the intensities of the process) as a result of the current score.

\citet{pollard} provided evidence for the existence of home advantage in soccer, and quantified it by the percentage of points gained from home games out of the total number of points. \citet{ClarkeNorman} suggested a more sophisticated model for the goal spread between the two teams, including team specific home advantage parameter, constructed relatively to the overall home advantage. The research in this area is still evolving, \citet{CarmThomas} used a large amount of data per game to construct and compare between different regression models for the home team and the away team goals and shots. They concluded that while attacking abilities are more important to the home team, defensive attributes are more important to the away team. According to them, this result could be explain by teams different game tactics home and away. \citet{HockeyJones} examined the home advantage in Hockey (a sport with many similarities to soccer). He specifically identified the home advantage for the scoring of a first goal, and its effect on the home team and the away team.

This paper aims at examining and quantifying the interaction between two random goal scoring times during a soccer match, using survival analysis methods. Survival analysis methods have long been used to investigate statistical phenomenons in disciplines other than the standard clinical trials framework. These methods may be applicable to any situation where the time until event occurrence is in question.  The survival analysis framework is used since our data including both censored and left truncated observations (censored and truncated by the end of the game, goal by the other team, etc.). The relevant survival analysis concepts are readjusted to fit the problem and a Cox model  is developed for the hazard function. Attributes such as time dependent covariates and a  frailty term are also being considered. The conclusions derived from the results are, in a nut shell, that a first goal occurrence could either expedite or impede the next goal scoring, depending on the time it was scored. Moreover, once a goal is scored, another goal scoring become more and more likely as the game progresses. Furthermore, and more interesting, the first goal effect is the same whether the goal was scored or conceded.

The rest of the paper is organized as follows.  In section \ref{smethods} the censoring mechanism, the possible models and the available data are presented;   Section \ref{sresults} includes the results which are later discussed in section \ref{sdiscu}.

\section{Methods}
\label{smethods}

We used data about  760 games which took place during the years 2008-2010 (two full seasons) in the Premier League, the major soccer league in England. Each game is formally 90 minutes long, divided into two halves. The goal times are accurate up to a minute (with the exception of goals scored in stoppage time, see section \ref{sdiscu}). Our background data was obtained from \textit{http://www.football-data.co.uk}. The goal scoring times were taken from \textit{http://www.soccerbot.com}.

Our interest was focused on the first two goals during a soccer match.  We believe that  the phenomenon of scoring a goal before any goal was scored is fundamentally different from scoring a goal when one goal has already been scored or conceded. Thus, from each game, two different observations are obtained:  \textit{FirstGoalTime} and   \textit{SecondGoalTime}. We consider a goal of the home team as the event in question and a goal of the away team as a censored observation. If no goals were scored by the end of the game, we have a censored observation at time 90. Since the second goal could not be observed until the first one took place, \textit{SecondGoalTime} is left truncated. The second goal could be either a first goal scored by the home team while a goal was conceded earlier, or a second goal scored by the home team. We emphasize that the relationship between the time of the first goal in the game and the time of the second goal is of our interest. One may target more goals in the game, although one would deal with heavier censoring and may need a much more complicated model and more data than probably possible (as more data would probably mean introducing much more heterogeneity). 

\begin{remark} We use the notion \textit{`FirstGoalTime'} for the dependent variable  - the time of the first goal of the home team, which is possibly censored. We will use the notation \textbf{`TimeOfFirstGoal'} for the independent variable - the time of the first goal in the game, regardless of which team scored the goal. This variable is used as an explanatory variable for the \textit{`SecondGoalTime'}.
\end{remark}

For example, in the game between the teams \emph{Newcastle United} and \emph{West Ham United}, which took place on January $10^{th}$, 2009, four goals were scored. \emph{Newcastle} (the home team) scored goals on the $19^{th}$ minute and the $78^{th}$ minute, while \emph{West Ham} (the away team) scored goals on the $29^{th}$ minute and the $55^{th}$ minute. The two observations obtained from this game are event in time 19, and an observation starting in time 19 and censored in time 29.

This paper is mainly aimed at examining the connection between the two first goals. In order to do so, we would want to control other effects that are common to both observations in each game. In particular, since a better team is more likely to score both goals. Past models (e.g., \citealp{DixonColes}) concentrated on the estimation of parameters such as home advantage, teams' attack and defence skills , and other possible effects. There are other important effects which would be hard to get from our data. For example, weather, player injuries, management problems, etc., which are not accessible by us, but were publicly known before the game.  We preferred, therefore, to use  one variable as a propensity score that summarizes all covariates  known prior to the game. \textbf{ProbWin} is the probability of the home team to win the game, based upon an average of the odds as determined by online gambling sites. We obtained the data from \textit{http://www.BetBrain.com}, which is a website designed to compare between different gambling odds for different bookmakers. Our data includes the average odds for at least 30 bookmakers per game. A preliminary analysis has shown that \textbf{ProbWin} is missing the difference between games played in different seasons, hence the variable \textbf{Season} is also considered.

\begin{table}[t]
\caption{Independent Variables}
\label{tab1}
\begin{tabular}{l|p{10cm}}
Variable & Description\\
\hline
ProbWin& Probability of a home team win\\
Season& Season 1:2009-10 0:2008-9\\
Goal& Indicator, goal already scored in the game 1:Yes 0:No*\\
TimeOfFirstGoal& Time of the first goal in the match*\\
FirstGoalTeam& Indicator, team scored the first goal 1:Home 0:Away*\\
TimeFromFirstGoal& Time passed since the first goal was scored*\\
\end{tabular}
\par\medskip\footnotesize
* These variables  are relevant only  for  the \textit{SecondGoalTime}.
\end{table}
See table \ref{tab1} for the definition of each variable. Note that the variable \textbf{TimeFromFirstGoal} is a time dependent variable.

Almost 10 percent of the games (74 out of 760) ended with no goals being scored. There are also 13 games in which a first goal was scored (either by the home team or the away team) in the $90^{th}$ minute. For those games we actually have only one available observation, the  \textit{FirstGoalTime}  at the $90^{th}$ minute (either as an event  or   censored by the end of the game). Therefore, the total number of observations is 1433, almost half of them (711)  were  censored. It should be also mentioned that about 46 percent (352 out of 760) of the \textit{FirstGoalTime} observations and 53 percent (359 out of 673) of \textit{SecondGoalTime} observations were censored.
The mean of \textbf{ProbWin}, based on the 1433 observations, was  0.448 (sd=0.187). The result was not substantially different when the calculation was based on the 760 games. Out of 673 first goals, 402 (60\%) were scored by the home team. The mean of \textbf{TimeOfFirstGoal},   the average of the time until the first goal, given that one was scored,  was 30.8 (sd=22.13). This is merely descriptive statistic, and not an estimate for \textit{FirstGoalTime} expectation.

The availability of explanatory variables, particularly time dependent variables, are among the advantages of the Cox  proportional hazard model (\citealp{Cox}). Assuming independent censoring, we model the hazard function by:
\newcounter{model}
\setcounter{model}{1}
\begin{equation*}
\tag{\Roman{model}}
\begin{split}
h_{ij}(t)&= h_0(t)\exp[\beta_1\mbox{ProbWin}_{i}+\beta_2\mbox{Season}_{i}
\\
&\hspace{3em}+I_{\{j=2\}}(\beta_3\mbox{Goal}_{i}+\beta_4\mbox{TimeOfFirstGoal}_{i}
\\&\hspace{4em}
+\beta_5\mbox{FirstGoalTeam}_{i}+\beta_6\mbox{TimeFromFirstGoal}_{i})],\\
\end{split}
\end{equation*}
where $i=1,...,760$ is the game number and $j=1,2$ is the goal number.

Using the notations $\tilde{t}_{ij}$ for the observed time and $\delta_{ij}$ that equals one only if the observed goal was scored by the home team, i.e., $\delta_{ij}=1$ if and only if $\tilde{t}_{ij}=t_{ij}$, where $t_{ij}$ is the uncensored (and not necessarily observed) goal time, the hazard function could be rewritten as
\begin{equation*}
\setcounter{model}{1}
\tag{\Roman{model}'}
\begin{split}
h_{ij}(t)&= h_0(t)\exp[\beta_1\mbox{ProbWin}_{i}+\beta_2\mbox{Season}_{i}+\\
&+I_{\{j=2\}}(\beta_3+\beta_4\tilde{t}_{i}+\beta_5{\delta_{i1}}+\beta_6(t-\tilde{t}_{i}))].\\
\end{split}
\end{equation*}
The indicator $I_{\{j=2\}}$ splits the hazard into two parts. The first part is common to both goals and related to all covariates that are known prior to the match. The second part is the effect of the first goal on the \textit{SecondGoalTime}.

This model distinguishes between different effects of the first goal; $\beta_3$ represents the basic difference between the two goals; $\beta_4$ represents the difference between early and late first goals; $\beta_5$ is used for the possible difference between a first goal scored and a first goal conceded; $\beta_6$ is a measure of decreasing or increasing effect of a goal as the game continues.

\begin{remark}
The independent censoring assumption should be clarified. We assume that the events of a goal for the home team and a goal for the away team are statistically independent. If, for example, the home team controls the game and creates chances for goal scoring, it might imply that an away team goal is seemingly less expected, but on the same time, the away team counter attacks are more likely and hence an away team goal is, on that score, more likely. Over all, given the conditions of the game, it seems reasonable to consider the two events as locally independent. It should however, be noted, that a goal scored by any of the teams, may effect the rest of the game, as will be in the center of the following discussion.
\end{remark}

Although model (\Roman{model}') includes all possible effects of the first goal occurrence, it might neglect some unobserved dependency between the goal times, which may be a result of game-specific unpredictable conditions, such as weather, crowd behavior during the game, incompetent referee etc.

We would consider the possibility of dependency by adding a frailty term i.e., a random effect, to the model. The hazard function for this model is
\begin{equation*}
\setcounter{model}{2}
\tag{\Roman{model}}
\begin{split}
h_{ij}(t)&= Z_ih_0(t)\exp[\beta_1\mbox{ProbWin}_{i}+\beta_2\mbox{Season}_{i}+\\
&+I_{\{j=2\}}(\beta_3+\beta_4\tilde{t}_{i}+\beta_5{\delta_{i1}}+\beta_6(t-\tilde{t}_{i}))]\\
\end{split}
\end{equation*}
where the $Z_i$'s are i.i.d. gamma distributed random variables with mean 1 and unknown variance $\theta$. \citet{Nielsen} and \citet{KleinEM} suggested an EM algorithm for the parameters estimation in this model.

In addition to models \setcounter{model}{1}(\Roman{model}) and \setcounter{model}{2}(\Roman{model}), we may consider simpler models. For example, we could consider the complete independence model,
\begin{equation*}
\setcounter{model}{3}
\tag{\Roman{model}}
\begin{split}
h_{ij}(t)&= h_0(t)\exp(\beta_1\mbox{ProbWin}_{i}+\beta_2\mbox{Season}_{i}),\\
\end{split}
\end{equation*}
meaning the hazard is the same for $j=1,2$. We treat this model as our null model.
\section{Results}
\label{sresults}

The data were analyzed using the package  `survival' for \textbf{R}  by Terry Therneau. The results of the full model, model \setcounter{model}{2}(\Roman{model}),  are presented  in table \ref{tab3}. The fixed covariates \textbf{ProbWin} and \textbf{Season} are highly significant and significant, respectively.  As for the \textit{SecondGoalTime} part of the hazard, only the variable \textbf{FirstGoalTeam} is undeniably not significant.
\newcolumntype{d}{D{.}{.}4}
\begin{table}[h!]
\caption{Results of Model \setcounter{model}{2}(\Roman{model})}
\label{tab3}
\begin{center}
\begin{tabular}{l|ddddd}
\hline
& \multicolumn1c{coef} & \multicolumn1c {exp(coef)} & \multicolumn1c {se(coef)} & \multicolumn1c {z} & \multicolumn1c {$p$} \\
\hline
ProbWin & 1.932 & 6.900 & 0.209 & 9.222 & <0.001 \\
Season & 0.150 & 1.162 & 0.074 & 2.012 & 0.044 \\
Goal & -0.329 & 0.720 & 0.200 & -1.649 & 0.099 \\
TimeOfFirstGoal & 0.012 & 1.012 & 0.004 & 2.995 & 0.003 \\
FirstGoalTeam & -0.054 & 0.948 & 0.118 & -0.455 & 0.649 \\
TimeFromFirstGoal & 0.008 & 1.008 & 0.004 & 2.005 & 0.045 \\
\hline
\end{tabular}
\end{center}
\end{table}

In order to test $H_0:\theta=0$ we use a likelihood ratio test. We get a chi-square statistic of 0.0003 on 1 degree of freedom ($p=0.986$) , meaning there is no evidence to include a frailty term in the model. These results were consistent for all models we have considered. Thus, we would include the covariates \textbf{ProbWin} and \textbf{Season} in the final model and we would disregard the variable \textbf{FirstGoalTeam} and the possibility of including a frailty term in the model from now on.

If we limit ourself to models in which all effects are at least weakly significant ($p<0.1$) we get two possible models,
\begin{equation*}
\setcounter{model}{4}
\tag{\Roman{model}}
\begin{split}
h_{ij}(t)&= h_0(t)\exp[\beta_1\mbox{ProbWin}_{i}+\beta_2\mbox{Season}_{i}+I_{\{j=2\}}\beta_4\tilde{t}_{i}]\\
\end{split}
\end{equation*}
and
\begin{equation*}
\setcounter{model}{5}
\tag{\Roman{model}}
\begin{split}
&h_{ij}(t)\\&= h_0(t)\exp[\beta_1\mbox{ProbWin}_{i}+\beta_2\mbox{Season}_{i}+I_{\{j=2\}}(\beta_3+\beta_4\tilde{t}_{i}+\beta_6(t-\tilde{t}_{i}))].\\
\end{split}
\end{equation*}
The results for these models are displayed in tables \ref{tab4} and \ref{tab5}. Note that both models include \textbf{TimeOfFirstGoal}, though the coefficient and the standard error estimates are twice larger for model \setcounter{model}{5}(\Roman{model}).

\begin{table}[t]
\caption{Results of Model \setcounter{model}{4}(\Roman{model})}
\label{tab4}
\begin{center}
\begin{tabular}{l|ddddd}
\hline
& \multicolumn1c{coef} & \multicolumn1c {exp(coef)} & \multicolumn1c {se(coef)} & \multicolumn1c {z} & \multicolumn1c {$p$} \\
\hline
ProbWin & 1.901 & 6.693 & 0.204 & 9.332 & <0.001 \\
Season & 0.150 & 1.162 & 0.074 & 2.028 & 0.043 \\
TimeOfFirstGoal & 0.006 & 1.006 & 0.002 & 2.479 & 0.013 \\
\hline
\end{tabular}
\end{center}
\end{table}
\begin{table}[t]
\caption{Results of Model \setcounter{model}{5}(\Roman{model})}
\label{tab5}
\begin{center}
\begin{tabular}{l|ddddd}
\hline
& \multicolumn1c{coef} & \multicolumn1c {exp(coef)} & \multicolumn1c {se(coef)} & \multicolumn1c {z} & \multicolumn1c {$p$} \\
\hline
ProbWin & 1.909 & 6.749 & 0.204 & 9.362 & <0.001 \\
Season & 0.150 & 1.162 & 0.074 & 2.017 & 0.044 \\
Goal & -0.363 & 0.695 & 0.184 & -1.974 & 0.048 \\
TimeOfFirstGoal & 0.012 & 1.012 & 0.004 & 2.992 & 0.003 \\
TimeFromFirstGoal & 0.008 & 1.008 & 0.004 & 2.019 & 0.043 \\
\hline
\end{tabular}
\end{center}
\end{table}
The three models are nested, and hence a likelihood ratio test could be used. Comparing model (\Roman{model}) and model \setcounter{model}{4}(\Roman{model}), separately, to model \setcounter{model}{3}(\Roman{model}) yields a $p$-value of 0.014 for both tests. By comparing models \setcounter{model}{5}(\Roman{model}) and \setcounter{model}{4}(\Roman{model}) a chi-square statistic of 4.58 on 2 df is obtained ($p$=0.101), meaning model \setcounter{model}{5}(\Roman{model}) should not be preferred over model \setcounter{model}{4}(\Roman{model}).

Note that model \setcounter{model}{5}(\Roman{model}) includes the time dependent term $(t-\tilde{t}_{i})$. We could also consider some other function $g(t,\tilde{t}_{i})$. The natural log function, $g(t,\tilde{t_{i1}})=\log(t-\tilde{t}_{i})$, seems to fit the data better since it preserves the significance of all covariates in the model and improves the likelihood of the model. The results for this model,
\begin{equation*}
\setcounter{model}{6}
(\Roman{model})
\begin{split}
h_{ij}(t)&= h_0(t)\exp[\beta_1\mbox{ProbWin}_{i}+\beta_2\mbox{Season}_{i}+\\
&+I_{\{j=2\}}(\beta_3+\beta_4\tilde{t}_{i}+\beta_6\log(t-\tilde{t}_{i}))],\\
\end{split}
\end{equation*}
are displayed in table \ref{tab6}. If we repeat the likelihood ratio test to compare between this model and model \setcounter{model}{4} (\Roman{model}), we get a 5.93 chi square statistic on 2 df, which yields a borderline $p$-value of 0.052.
\begin{table}[t]
\caption{Results of Model \setcounter{model}{6}(\Roman{model})}
\label{tab6}
\begin{center}
\begin{tabular}{l|ddddd}
\hline
& \multicolumn1c{coef} & \multicolumn1c {exp(coef)} & \multicolumn1c {se(coef)} & \multicolumn1c {z} & \multicolumn1c {$p$} \\\hline
ProbWin & 1.915 & 6.788 & 0.204 & 9.384 & <0.001 \\
Season & 0.151 & 1.163 & 0.074 & 2.030 & 0.042 \\
Goal & -0.594 & 0.552 & 0.257 & -2.313 & 0.021 \\
TimeOfFirstGoal & 0.011 & 1.011 & 0.004 & 3.117 & 0.002 \\
$\log$TimeFromFirstGoal & 0.160 & 1.173 & 0.070 & 2.280 & 0.023 \\
\hline
\end{tabular}
\end{center}
\end{table}

In order to choose the most suitable model, we  have tried  to fit models \setcounter{model}{6}(\Roman{model}) and \setcounter{model}{4}(\Roman{model}) using only \textit{SecondGoalTime} observations. Obviously, the variable \textbf{Goal} was  eliminated from this study. All together we have 673 observations. The surprising result is that none of these models fit well, i.e., the effects related to the first goal are not significant.

Next,  a stratified model with no interactions between the two goals was fitted to the data. That is, \begin{equation*}
\setcounter{model}{7}
(\Roman{model})
\begin{split}
h_{ij}(t)&= h_{0j}(t)\exp(\beta_1\mbox{ProbWin}_{i}+\beta_2\mbox{Season}_{i}),\\
\end{split}
\end{equation*}
and looking at the differences between the baseline cumulative hazard functions $\hat{H}_{01}(t)$ and $\hat{H}_{02}(t)$  may reveal which model is more adequate.

\begin{figure}[t]
\caption{Cumulative Baseline Hazard for Model (\Roman{model})}
\label{fig6}
\begin{center}
\includegraphics[width=0.7\textwidth]{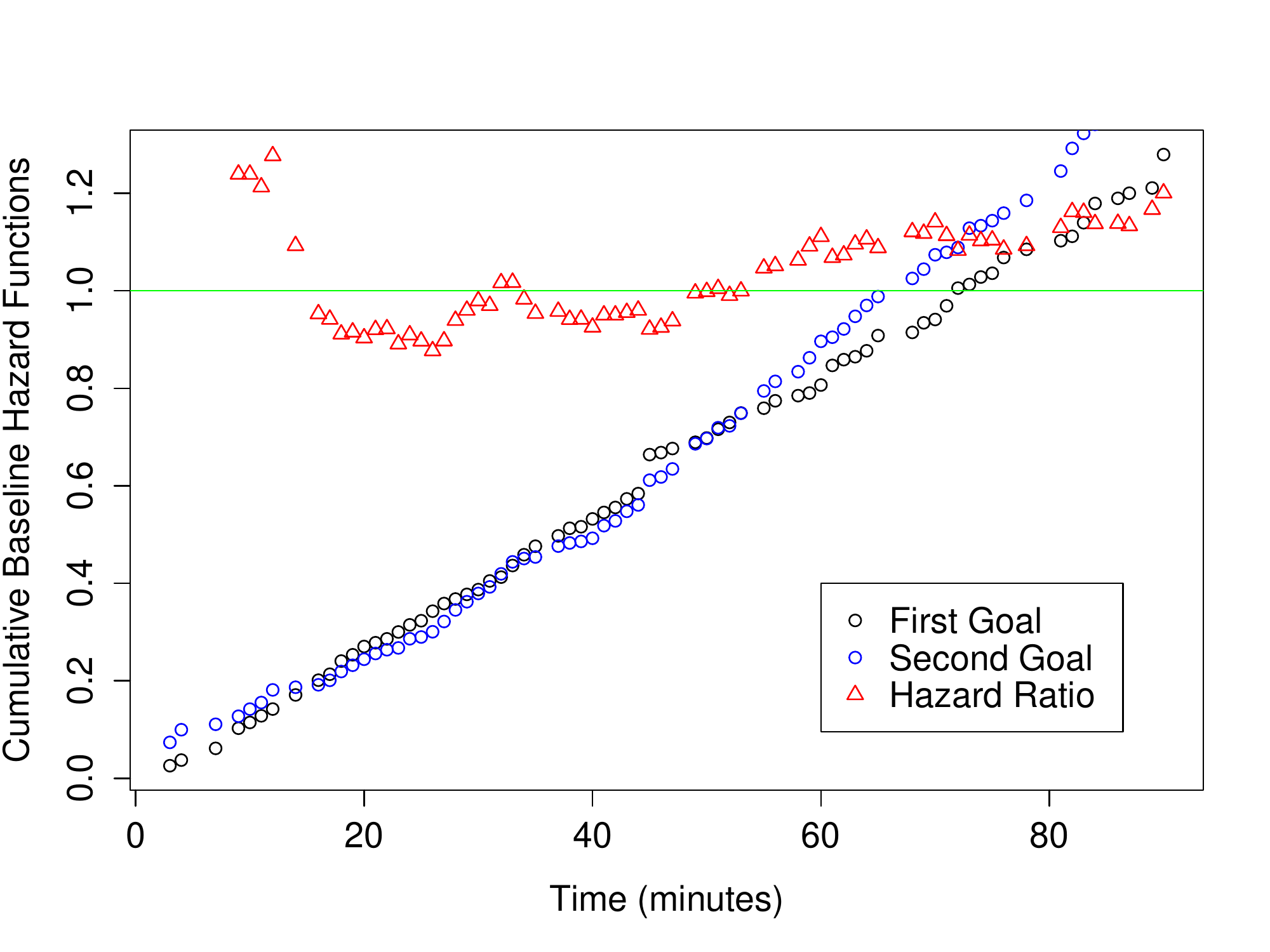}
\end{center}
\end{figure}

Figure \ref{fig6} presents these baseline cumulative hazards. The baseline hazard function for the \textit{FirstGoalTime} seems to be linear. The \textit{SecondGoalTime} baseline hazard, although close to linear, behaves differently at the edges. The hazard between the $20^{th}$ and the $70^{th}$ minutes is similar to the \textit{FirstGoalTime} hazard. Looking at the hazard ratio reveals that as time goes by, a second goal is more and more likely to be scored in comparison to  a first goal. These findings strengthen the need of a time dependent term for the \textit{SecondGoalTime} hazard.

The hazard for the First 15 minutes behaves differently. A goal is more likely to be scored if one has already been scored. Models \setcounter{model}{4}(\Roman{model}) and \setcounter{model}{6}(\Roman{model}) miss this phenomenon. We have tried to include appropriate effect in our models, but it was rejected. It should be noted that only 29 out of the 673 \textit{SecondGoalTime} observations were in the first 15 minutes and therefore it is harder to model them appropriately.

We would summarize the model choice discussion by stating that the first goal may effect the \textit{SecondGoalTime} by a few small effects. When combining them together they are significant but each of them separately is not strong enough to be found significance. Moreover, we have presented evidence to time dependent changes for the \textit{SecondGoalTime} hazard. Therefore, we suggest that model (\setcounter{model}{6}\Roman{model}) is more appropriate.

\section{Discussion}
\label{sdiscu}

\begin{figure}[h!]
\caption{Model (\setcounter{model}{6}\Roman{model}) Survival Curves for Different Values of \textbf{ProbWin}}
\label{survprob}
\begin{center}
\includegraphics[width=0.7\textwidth]{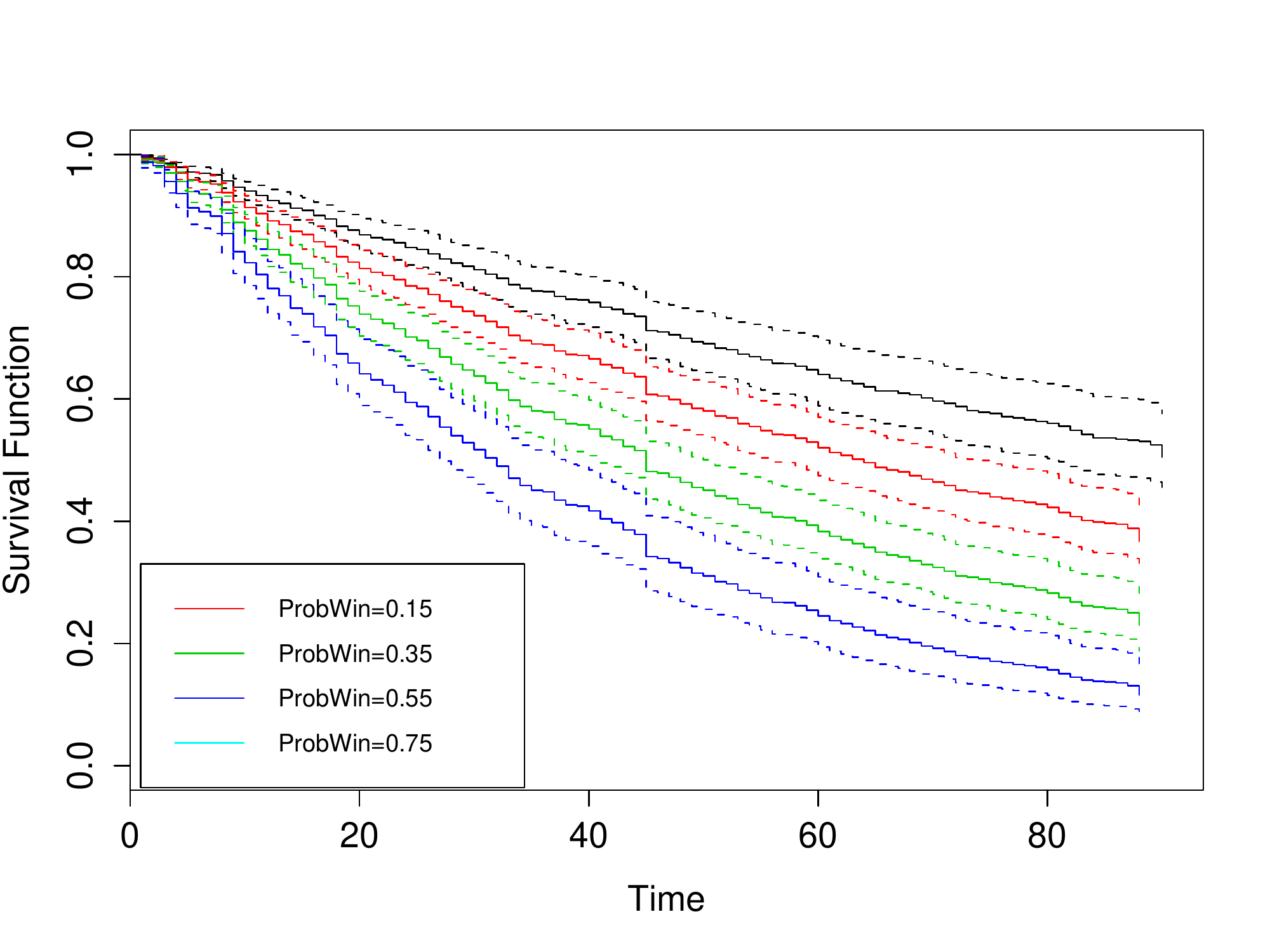}
\end{center}
\end{figure}

The estimates for the different effects included in the chosen model (presented in table \ref{tab6}) should be interpreted more accurately. \textbf{ProbWin} positively effects on the hazard. If the difference between \textbf{ProbWin} of two home teams is 0.1, then their hazard ratio is $e^{0.1\times 1.915}=1.21$. This result was expected since generally, better teams score goals faster. Figure \ref{survprob} presents \textit{FirstGoalTime} survival curves for four theoretic teams with the win probabilities of 0.15, 0.35, 0.55 and 0.75 (all in the same season). The dashed lines are $95 \%$ confidence intervals for the survival function.

When the first goal is scored, the immediate effect on the hazard could be either positive or negative. The basic effect reduces the hazard almost by half ($e^{-0.59}=0.552$). However, there is a positive effect of \textbf{TimeOfFirstGoal}. Therefore, the immediate multiplicative effect on the hazard, in comparison to a situation where no goal had been scored yet, is $e^{-0.59+0.011\textbf{TimeOfFirstGoal}}$.
We conclude that there is no immediate effect when a goal is scored in the $52^{nd}$ minute. If a goal is scored before $52^{nd}$ minute the effect is negative, and if it is scored after that the effect is positive.

\begin{figure}[h!]
\caption{Model (\setcounter{model}{6}\Roman{model})  Survival Curves for Different \textbf{TimeOfFirstGoal} }
\label{survgoals}
\begin{center}
\includegraphics[width=0.7\textwidth]{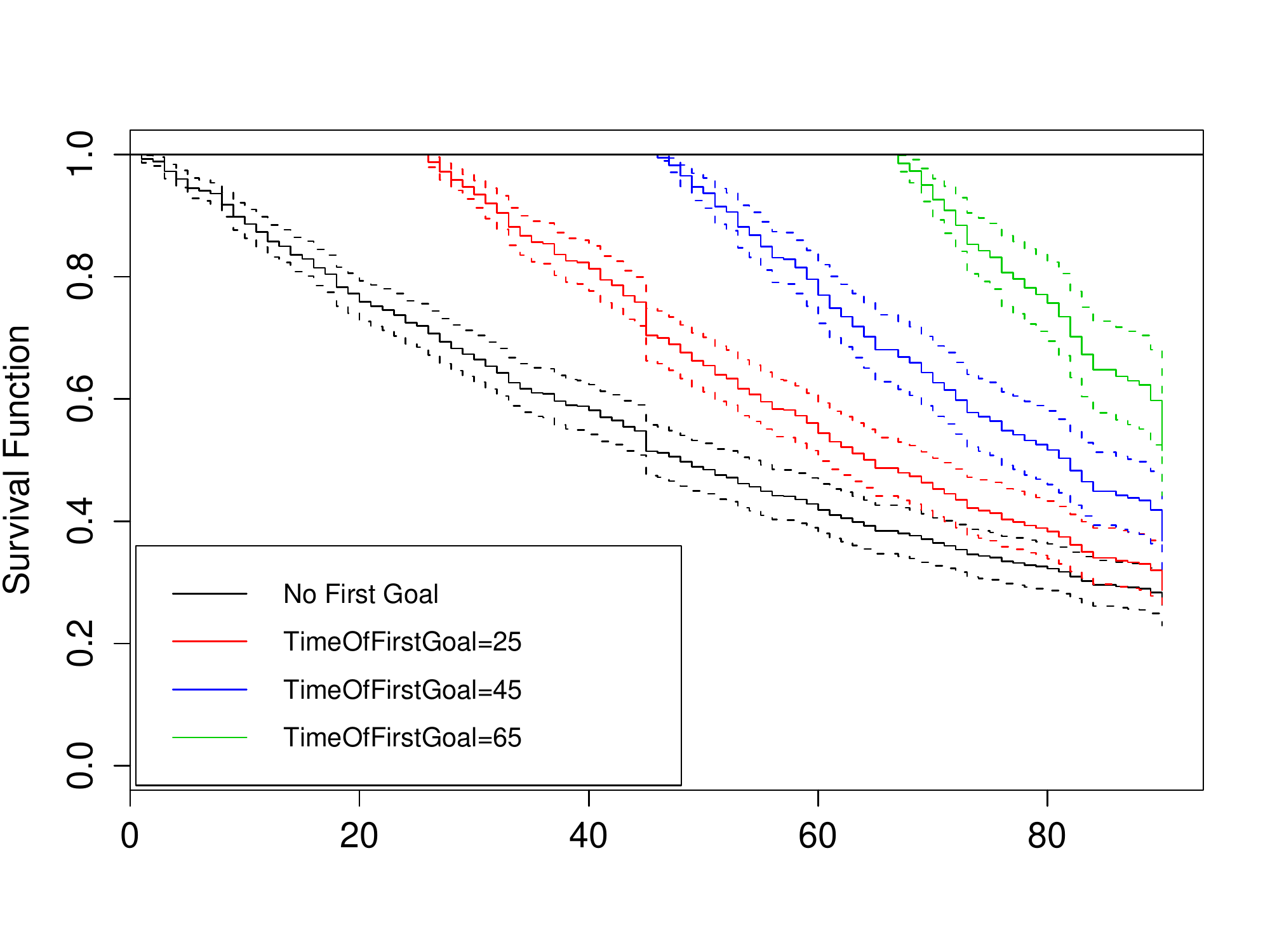}
\end{center}
\end{figure}

There is also a time dependent effect in this model. 5 minutes after \textbf{TimeOfFirstGoal}, the hazard is $e^{0.1597\times\log(5)}=1.293$ times greater than the hazard when the goal was scored. 30 minutes after \textbf{TimeOfFirstGoal} the hazard is 1.72 times greater. In figure \ref{survgoals} survival curves for possible theoretic situations are presented. The black line is the estimated survival curve of \textit{FirstGoalTime} for a team with 0.5 probability to win. The other lines are survival curves for \textit{SecondGoalTime}, for different \textbf{TimeOfFirstGoal}, all with $95 \%$ confidence interval.

Once a goal is scored, another goal is less likely to be scored, compared to the situation where a goal was not scored yet. As the game nears its end, this difference shrinks, and in the last few minutes of the game, goal scoring is more expected. This phenomenon could be explained in the following way:
when a game begins, both teams play according to pre-determinant tactics. If no goal is scored or an early goal is scored, both teams do not make substantial changes in their tactics, because there is still enough time to score a goal. As the game progresses, the need to score a goal increases and hence tactics become adventurous. In addition, a coach may consider substituting players. These changes could lead to a goal scoring, for both sides.

As stated before, the impact of the first goal could be divided into a few small effects. However, there was no evidence to support including an independent effect for \textbf{FirstGoalTeam}. This implies that whether a team is leading or behind by a goal, their chances of scoring a goal remain the same. This result could be explained by the way teams interact during a game. If one team attacks more aggressively, it also weakens its defence and could lead to a goal scoring for each side.

\begin{figure}[h!]
\caption{Cumulative Baseline Hazards for Model (\setcounter{model}{6}\Roman{model})}
\label{cumbasehazmodel6}
\begin{center}
\includegraphics[width=0.7\textwidth]{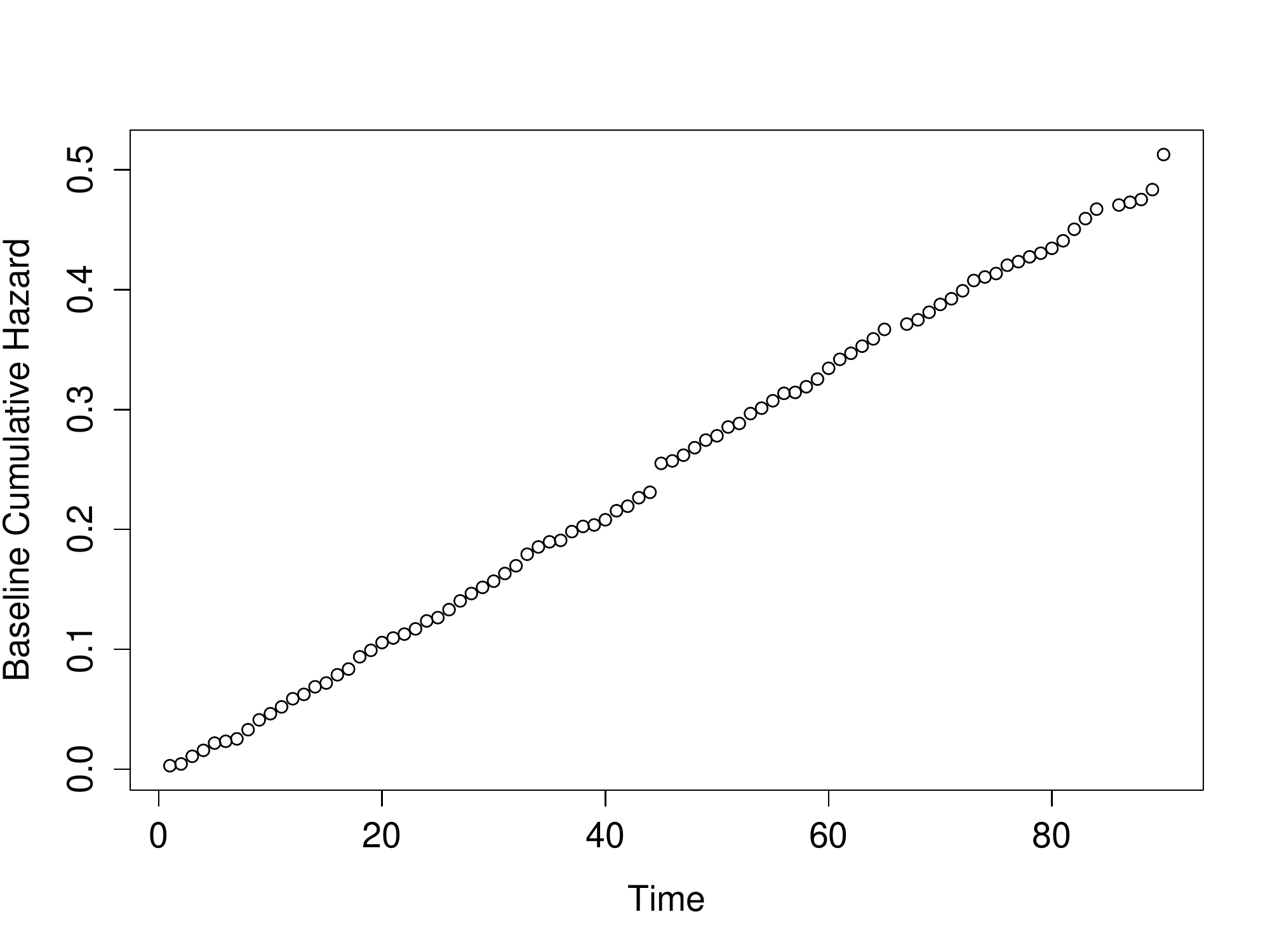}
\end{center}
\end{figure}

The baseline cumulative hazard estimator, $\hat{H}_0(t)$, is also of interest. The baseline hazard is the hazard for a subject with covariates all equal to zero. Although our data does not include such a subject (all teams have positive probability to win the game), we could be interested in the shape of the baseline hazard. $\hat{H}_0(t)$ is presented in figure \ref{cumbasehazmodel6}. It seems to be almost linear. This implies an exponential distribution for the goal time, as generally assumed. However, Since the \textit{SecondGoalTime} hazard behaves differently than the \textit{FirstGoalTime} hazard, due to the time dependent effect, this conclusion is admissible only for the first goal time.

The jumps in the $45^{th}$ minute and the $90^{th}$ minute, which appear in other figures as well, are due to a special property of soccer data. In the end of each half the referee adds a few minutes called ``stoppage time'', also called ``injury time''. The referee decides at the end of each half how much time to add in the end of every half to cover for the time the game was not played as a result of injuries, substitutions, etc. It is typically a few minutes. The time added does not count as part of the 90 minutes and goals scored during this time are recorded as if they were scored in the last minute of the appropriate half. Therefore, more goals are scored in these minutes than any other minute.

We conclude from the absence of evidence to include a frailty term in the model, that there is no positive dependency between the two goals as a result of the game-specific conditions. The frailty term was found non-significant even when we considered it for a model with no goal interactions (model \setcounter{model}{3}(\Roman{model})). Results of other models have shown that dependency might be negative, i.e., a late first goal is evidence for a quick second goal. The frailty scheme assumes positive dependency and hence is not adequate to our purposes.

\begin{figure}[h!]
\caption{Deviance Residuals for Model (\Roman{model})}
\label{devfig}
\begin{center}
\subfloat[Part 1][Varaible ProbWin]{\includegraphics[width=0.48\textwidth]{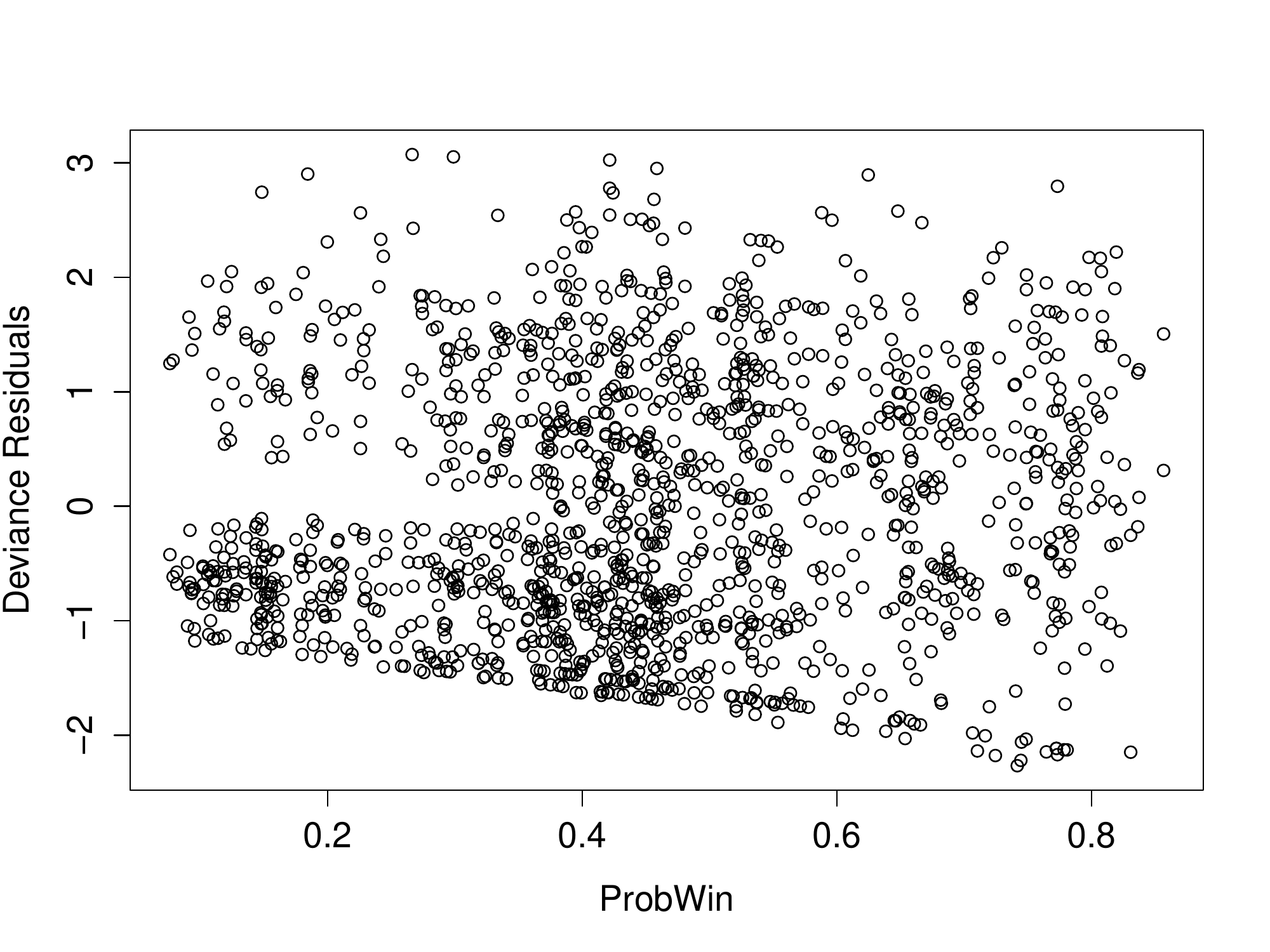} \label{devfig1}}
\subfloat[Part 2][Varaible TimeOfFirstGoal]{\includegraphics[width=0.48\textwidth]{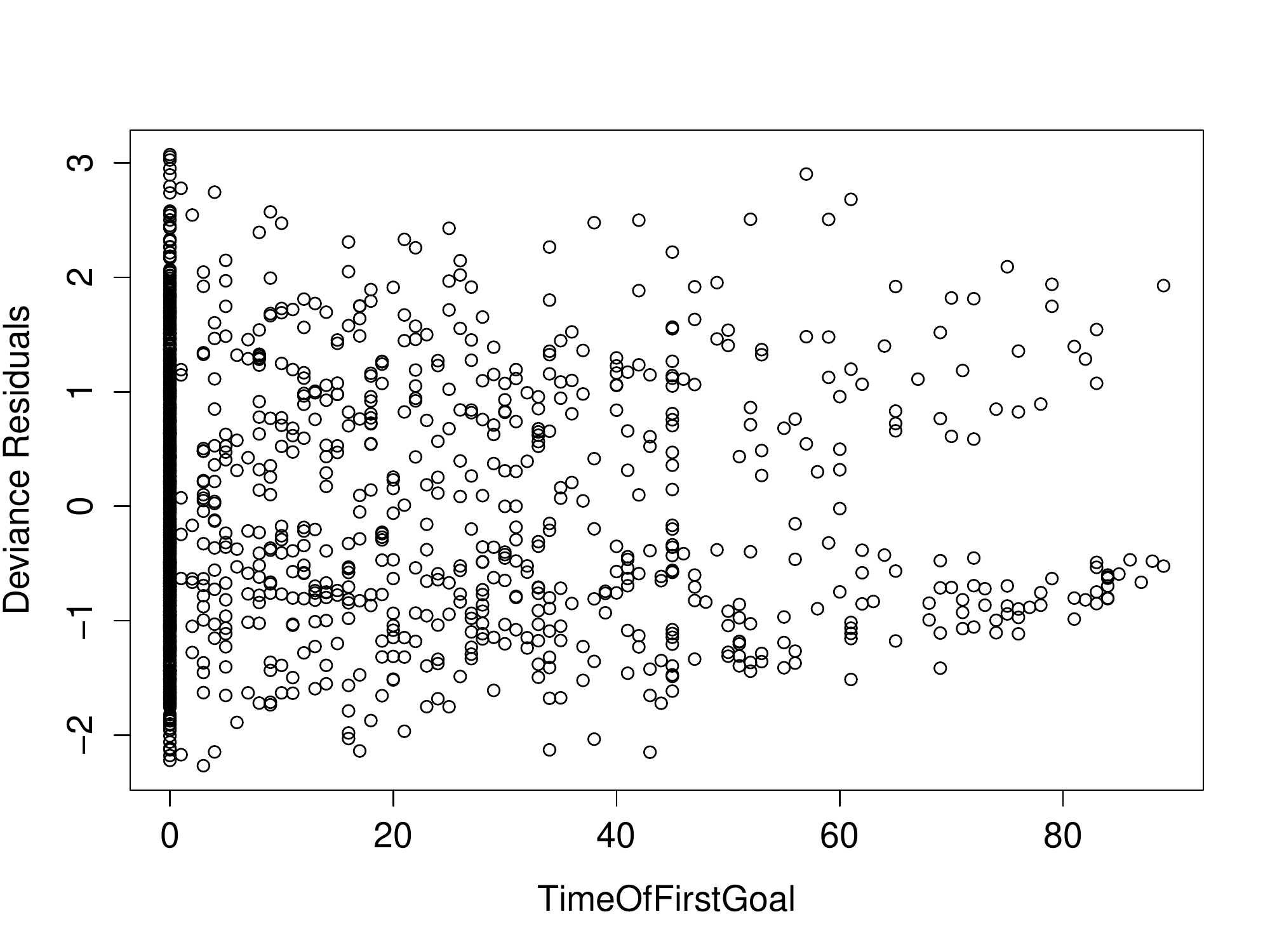} \label{devfig2}}
\end{center}
\end{figure}

Although our model seems to be appropriate, it has its own shortcomings. We use the deviance residuals as suggested in \citet{therneaumartingale} in order to detect model errors. For example, goalless matches, especially ones where the home team is stronger. Additionally, there are more positive outliers than negative ones, as can be seen in figure \ref{devfig}. Therefore, our model's major errors are due to surprising fast goals. It was mentioned earlier that the model is inaccurate for the first 15 minutes of the game. We conclude that these aspects of the game may be more random than others.\setcounter{model}{6}
\pagebreak
\bibliographystyle{plainnat}
\bibliography{FirstGoalEffect}
\end{document}